\documentclass[aps,prb,showpacs,amsmath,superscriptaddress,twocolumn,floatfix,nobalancelastpage]{revtex4-1}

\usepackage{hyperref}
\usepackage[all]{hypcap}
\usepackage{amsmath}
\usepackage{amsfonts}
\usepackage{bm}
\usepackage{graphicx}
\usepackage{grffile} % to handle dots in graphics filnames
\usepackage{color}
\usepackage{esint}
\usepackage{textpos}
\usepackage[usenames,dvipsnames]{xcolor}
\newcommand{\pa}{\partial}

\newcommand{\vphi}{\varphi}
\newcommand{\ve}{\varepsilon}

\newcommand{\Vect}[1]{\mbox{\boldmath$#1$}}

\def\widebar{\accentset{{\cc@style\underline{\mskip10mu}}}}

\begin{document}
%\begin{CJK*}{SJIS}{}
%\preprint{APS/123-QED}

%%%%%%%%%%%%%%%%%%%%%%%%%%%%%%%%%%%%%%%%%%%%%%%%%%%%%%%%%%%%%%%%
%%%%%%%%%%%%%%%%          TITLE         %%%%%%%%%%%%%%%%%%%%%%%%%%%%%%%%%%%%%%%
%%%%%%%%%%%%%%%%%%%%%%%%%%%%%%%%%%%%%%%%%%%%%%%%%%%%%%%%%%%%%%%%
\title{
Heterodyne Hall Effect in a Two Dimensional Electron Gas
}

%%%%%%%%%%%%%%%%%%%%%%%%%%%%%%%%%%%%%%%%%%%%%%%%%%%%%%%%%%%%%%%%
%%%%%%%%%%%%%%%%          AUTHORS       %%%%%%%%%%%%%%%%%%%%%%%%%%%%%%%%%%%%%
%%%%%%%%%%%%%%%%%%%%%%%%%%%%%%%%%%%%%%%%%%%%%%%%%%%%%%%%%%%%%%%%

\author{Takashi~Oka} 
\affiliation{ Max Planck Institute for the Physics of Complex Systems,
N\"othnitzer Str. 38, D-01187, Dresden, Germany}
\affiliation{
Max Planck Institute for Chemical Physics of Solids,
N\"othnitzer Str. 40, D-01187, Dresden, Germany}
\author{Leda~Bucciantini}
\affiliation{ Max Planck Institute for the Physics of Complex Systems,
N\"othnitzer Str. 38, D-01187, Dresden, Germany}
\affiliation{
Max Planck Institute for Chemical Physics of Solids,
N\"othnitzer Str. 40, D-01187, Dresden, Germany}

\date{\today}
\begin{abstract}
\noindent 
 We study the hitherto un-addressed phenomenon of Quantum Hall Effect with a magnetic and electric field  oscillating in time with resonant frequencies. This phenomenon realizes an example of heterodyne device with the magnetic field acting as a driving and is analyzed in detail in its classical and quantum  versions using Floquet theory.  A bulk current flowing perpendicularly to the applied electric field is found, with a frequency shifted by integer multiples of the driving frequency. When the ratio of the cyclotron and driving frequency takes special values, the electron's classical trajectory forms a loop and the effective mass diverges, while in the quantum case we find an analogue of the Landau quantization. Possible realization using metamaterial plasmonics is discussed. 
 %Keywords: Floquet theory, Quantum Hall Effect, heterodyne, Landau quantization, plasmonics
\end{abstract}

\maketitle
%\end{CJK*}
%\section{ Introduction}

%%%%%%%%%%%%%%%%%%%%%%%%%%%%%%%%%%%%%%%%%%%%%%%%%%%%%%%%%%%%%%%%
%%%%%%%%%%%%%%%%         Introduction      %%%%%%%%%%%%%%%%%%%%%%%%%%%%%%%%%%
%%%%%%%%%%%%%%%%%%%%%%%%%%%%%%%%%%%%%%%%%%%%%%%%%%%%%%%%%%%%%%%%

\section{Introduction}\label{uno}

%%%%%%%%%%%%%%%%%%%%%%%%%%%%%%%%%%%%%%%%%%%%%
\begin{figure}[htb]
\centering 
\includegraphics[width=0.5\textwidth]{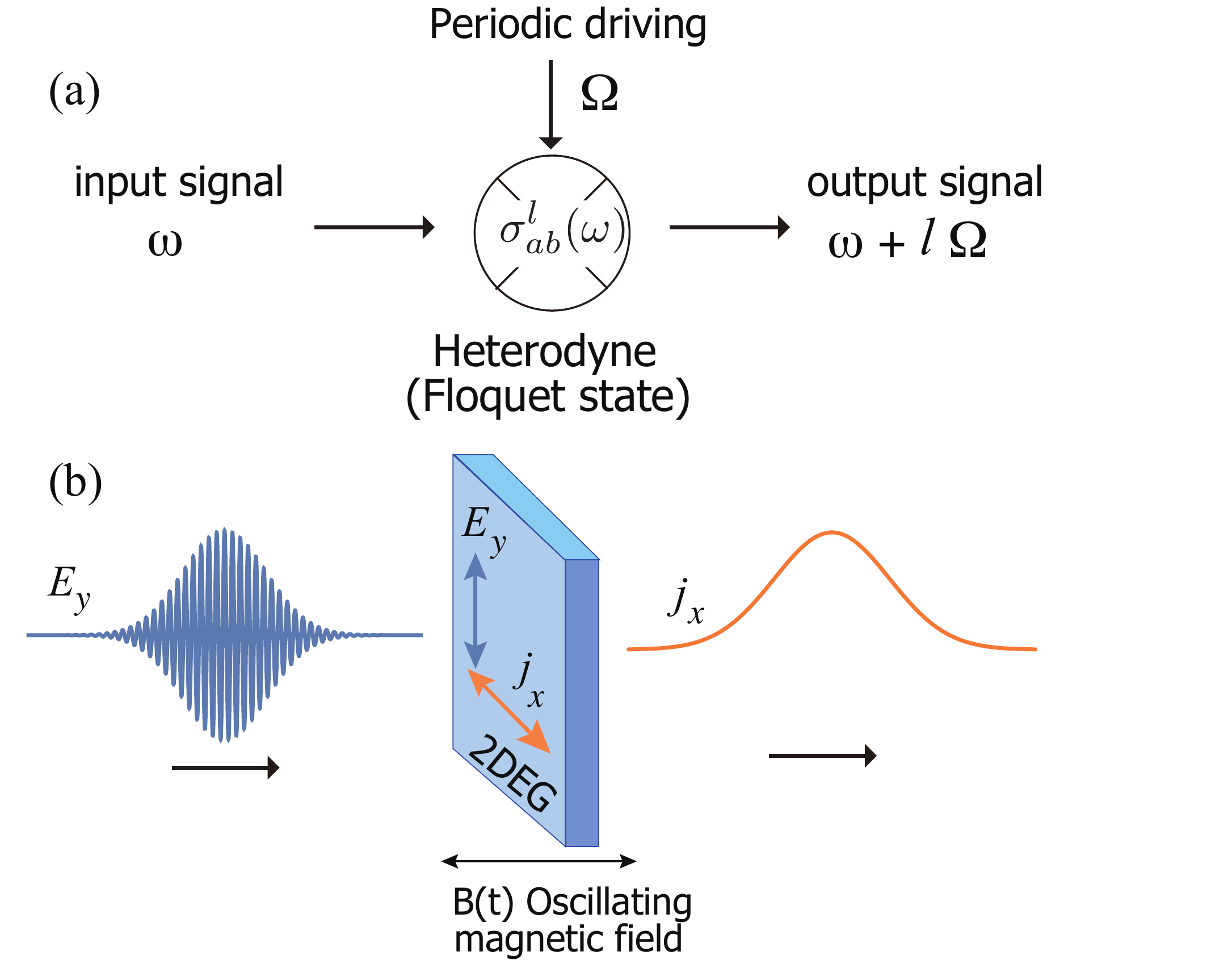}
\caption{(Color online) (a) 
A heterodyne mixes the frequency $\omega$ of the input signal  with its driving frequency $\Omega$. The output  is a superposition of signals with frequencies $\omega + l \Omega$, $ l \in \mathbb{Z}$. (b) Model under study: an input electric field directed along $y$  and a magnetic field oscillating in time along $z$, acting on  a two dimensional electron gas (2DEG): this yields an electric current along $x$, with a frequency different from the one of the input electric field.
In this example, the input is a single mode with an envelope function, and the output is the zero frequency envelope, 
which are related by the coefficient $\sigma^{-1}_{xy}(\Omega)$.%(=\sigma^{01}_{xy})$.
}
\label{fig:heterodyne}
\end{figure}
%%%%%%%%%%%%%%%%%%%%%%%%%%%%%%%%%%%%%%%%%%%%

Quantum Hall Effect (QHE) is one of the deepest phenomena in condensed matter physics. 
When a static electric field is applied to a quantum Hall state, a current perpendicular to the field 
is induced, and their linear relation $j_x=\sigma^HE_y$ is given by the Hall conductivity 
$\sigma^{H}=\frac{e^2}{h}\nu$ \cite{Klitzing80}.  
In Integer Quantum Hall Effect (IQHE), the factor $\nu$ is strictly an integer and was related to a
topological index, the 1st Chern number, by Thouless, Kohmoto, 
Nightingale and den Nijs \cite{TKNN}. 
The process is dissipationless because the current is perpendicular to the field and no Joule heating takes place. 

Here, we report an extension of this concept to the physically interesting case when the magnetic  and  electric fields are time dependent  with resonant frequencies. This realizes an example of heterodyne response, which is an ubiquitous  technique in today's electronics with various usages such as high precision optical detection \cite{maznev1998optical,Drever1983,Lenth:83}.
Heterodyne (frequency mixer) is an electronic device that mixes frequencies of oscillating signals through a nonlinear process [Fig. \ref{fig:heterodyne}(a)]. 
It is periodically driven by a ``local oscillator'' with  frequency $\Omega$, and 
integer multiples of $\Omega$ are added or subtracted to the 
frequency $\omega$ of the input signal. Here we will be interested in studying a heterodyne system where the driving oscillator  is the magnetic field  while the input signal  is an electric field [Fig. \ref{fig:heterodyne}(b)].

An important example of periodically driven systems is the zero resistance state that occurs 
in  a 2DEG driven by microwaves in a semiconductor heterostructure
in weak magnetic fields \cite{Mani2002} (reviewed in ref.\cite{MirlinRMP12}). 
More recently,  periodically driven lattice systems
are attracting interest \cite{Oka09,Kitagawa2011,Lindner2011,Rudner13} 
as a way to realize  a topological Chern insulator \cite{Haldane1988}, which was recently
confirmed experimentally \cite{Rechtsman2013,Jotzu2014}.
However,  we stress that these examples focused on the  response of the system to a static electric field, 
and the heterodyne response still waits for detailed investigation.  In this paper, we set out to fill this gap and  develop a theory for heterodyne response by studying the  conductivity of a 2DEG confined in the $xy$-plane subject to a $z$-directed  magnetic field
\begin{equation}
B_z(t)=B\cos\Omega t,
\label{eq:Bz}
\end{equation}
with an oscillating electric field [see  Fig. \ref{fig:heterodyne}(b)]. We will focus on the strong nonlinear effects introduced when the frequencies of the driving and of the electric field are resonant, i.e. when $\omega = n  \Omega$, $ n \in \mathbb{N}$.

The paper is organized as follows.  In Sec. \ref{due} and \ref{tre} we develop a theory for this system at the classical and quantum level respectively, while in Sec. \ref{quattro} we summarize our results and discuss open problems. 
 
%%%%%%%%%%%%%%%%%%%%%%%%%%%%%%%%%%%%%%%%%%%%
 \section{Classical case}\label{due}
%%%%%%%%%%%%%%%%%%%%%%%%%%%%%%%%%%%%%%%%%%%%

In this section we study the response of a classical 2DEG to a time oscillating weak electric field in the presence of an oscillating magnetic field and  compute the conductivity tensor, that we call heterodyne conductivity. The heterodyne conductivity $\sigma_{ab}^{m,n}$, introduced here for both classical and quantum cases, is  a four index tensor implicitly defined by the linear relation that holds between the electric current density $j_a(m\Omega)$ of the output signal with frequency $m \Omega$, flowing along the $a$-direction  ($a,b=x,y,z$) and  the (weak) electric field $E_b^n$ along the  $b$-direction with frequency $n \Omega$.

Given an  electric signal $E_b(t)=E_b(\omega) e^{-i\omega t}$ along a direction $b$ with frequency $\omega$,  the output current generated from the heterodyne along a direction $a$ 
can be expanded in modes with frequencies $\omega+l\Omega$, with $l$ a generic integer,  as $j_a(t)=\sum_l j_a(\omega+l\Omega)e^{-i(\omega+l\Omega) t}$. Then the linear relation 
\begin{equation}
j_a(\omega+l\Omega)=\sum_{b}\tilde\sigma_{ab}^{l}(\omega)E_b(\omega)
\label{eq:heterodyneresponse}
\end{equation}
holds as long as the field is weak and  defines the conductivity  $\tilde\sigma_{ab}^{l}(\omega)$. When $\omega=n\Omega$, with $n \in \mathbb{N}$, defining $E_b(n \Omega) \equiv E_b^n$ and $l+n=m$, (\ref{eq:heterodyneresponse}) can be rewritten as
\begin{equation}
j_a(m\Omega)=\sum_{b}\tilde\sigma_{ab}^{m-n}(n \Omega)E_b^n.
\label{eq:heterodyn0}
\end{equation}
Defining $\tilde\sigma_{ab}^{m-n}(n \Omega)\equiv \sigma_{ab}^{m,n}$, so that the upper left index labels the component of the outgoing current while the upper right index the  component of input electric field, then (\ref{eq:heterodyn0}) gives 
\begin{equation}
j_a(m\Omega)=\sum_{b}\sum_{n=-\infty}^\infty\sigma_{ab}^{m,n} E_b^n.
\label{eq:heterodyn}
\end{equation}
More explicitly, the heterodyne conductivity $\sigma^{m,n}_{ab}$ is obtained inverse Fourier transforming  (\ref{eq:heterodyn})
\begin{equation}
\sigma^{m,n}_{ab}=\lim_{t_0 \to \infty} \frac{1}{t_0 E_b^n}\int_0^{t_0} dt \;e^{i m \Omega t} j_a(t)
\end{equation}
with  $j_a(t)=e n_e v_a(t) $. 

The current density  $j_a(t)$ is related to the electron's velocity  $v_a(t)$ by the  relation $j_a(t)=n_e e v_a(t)$, with  $ e<0$  the electron's charge and $n_e$ the electron's density; the velocity $v_a(t)$  can be derived from the solution of the classical equation of motion 
\begin{equation}\label{eomgen}
m_e\left(\frac{d}{dt}+\eta\right)\Vect{ v}(t)=e\left(\Vect{E}+\frac{1}{c}\Vect{v}(t)\times\Vect{B}(t)\right),
\end{equation}
where $m_e$ is the electron's mass while $\eta$ is a small phenomenological scattering parameter  necessary for the convergence of the particle's trajectory in electric fields; $\Vect{B}(t)= B_z(t) \hat {\bf{z}}$ is the oscillating magnetic field (\ref{eq:Bz}), $\Vect{E}$ is the (infinitesimal) applied  electric field. We note that we have neglected the electric field emerging from the time dependent magnetic field, which will be recovered in the quantum case.  Given the rotational invariance of the system, we arbitrarily fix the direction of the  electric field  as the $y$-direction and restrict our analysis to $\{n,m\}=0,1$, with $E_y(t)=E_y^0$ ($n=0$)  and $E_y(t)=E_y^1 \cos(\Omega t)$ ($n=1$). The behavior of the particle strongly depends on the ratio
\begin{equation}
r=\frac{\omega_c}{\Omega},
\end{equation}
with $\omega_c=|e|B/m_ec$ the cyclotron frequency.

The formulas for the heterodyne conductivities can be derived as follows.  Defining
\begin{equation}
v(t)=v_x(t)+i v_y(t),
\end{equation}
the equation of motion (\ref{eomgen}) for  ${\bf{E}}=E_y^0 \hat {\bf{y}}$ becomes
\begin{equation}
\dot{v}(t)=i \frac{e E_y^0}{m_e} +v(t)\left(- i \frac{e B_z(t)}{m_e c} -\eta\right),
\end{equation}
whose solution is 
\begin{equation}\label{s11}
v(t)=i \frac{e E_y^0}{m_e} e^{-\eta t + i r \sin(\Omega t)} \int_0^t ds \;  e^{\eta s- i r \sin(\Omega s) }.
\end{equation}
 The results for the conductivities are thus
\begin{eqnarray}
&& \sigma_{x y}^{0,0}+i \sigma_{y y}^{0,0}=i\frac{e^2 n_e}{m_e} \sum_{n=-\infty}^\infty\frac{J_n(r)^2}{\eta-i \Omega n},\nonumber\\
&& \sigma_{x y}^{1,0}+i \sigma_{y y}^{1,0}=i \frac{e^2 n_e}{m_e}\sum_{n=-\infty}^\infty\frac{J_n(r)  J_{1-n}(r) (-1)^n}{\eta+i \Omega (1-n)}\nonumber,
\end{eqnarray}
where $J_n(r)$ is the $n$-th  Bessel function of the first kind.  From the former of these equations we derive  
\begin{equation}
\sigma_{x y}^{0,0}=0, \quad \quad \sigma_{y y}^{0,0}=\frac{e^2 n_e}{m_e \eta} J_0(r)^2.
\end{equation}
When applying an oscillating electric field along the $y$-direction 
with ${\bf{E}}(t)=E_y^1 \cos(\Omega t) \hat {\bf{y}}$, the solution for $v(t)$ is
\begin{equation}
v(t)=i \frac{e E_y^1}{m_e} e^{-\eta t + i r \sin(\Omega t)} \int_0^t ds \;  e^{\eta s- i r \sin(\Omega s) } \cos(\Omega s),\nonumber
\end{equation}
and, as a result, we get
\begin{equation}
\sigma_{x y}^{0,1}+i \sigma_{y y}^{0,1}=i \frac{e^2 n_e}{m_e}  \sum_{n=-\infty}^\infty\frac{J_n(x) J_{n-1}(x)}{\eta +i \Omega n}.
\end{equation}

%%%%%%%%%%%%%%%%%%%%%%%%%%%%%%%%%%%%%%%%%%%%%
\begin{figure}[htb]
\centering 
\includegraphics[width=.5\textwidth]{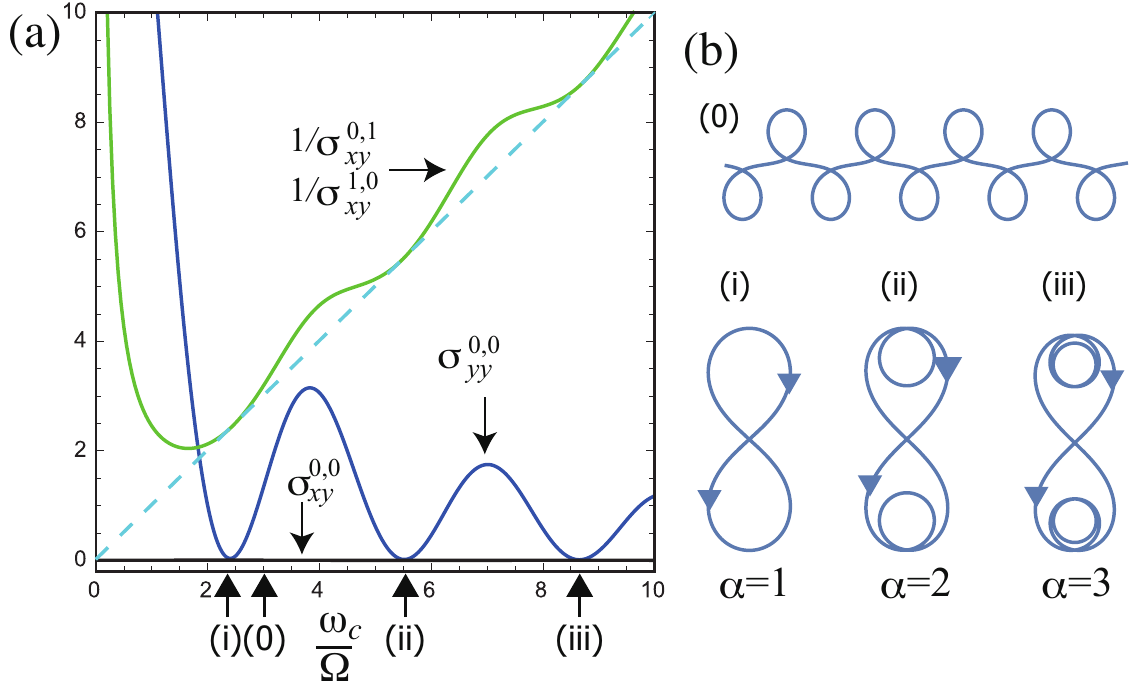}
\caption{(Color online) 
(a) Static and heterodyne conductivities  of a classical particle in an oscillating magnetic field. The dashed  line represents the classical result for the resistivity in a static magnetic field. The parameter $\eta$ is 0.05, $n_e=m_e=|e|=1$. $r=r_\alpha^{\rm cl}$ are identified as the zeroes of the  Bessel function $J_0(r)$, explicitly $r^{\rm cl}=\{2.40,5.52,8.67, \dots\}$.   
(b) Trajectories of the charged particle for different values of $r$, in zero electric field. 
(0) For general values, the particle makes open detours. 
(i-iii) For  $r_1^{\rm cl}=2.40$ (i), $r_2^{\rm cl}=5.52$ (ii) and $r_3^{\rm cl}=8.67$ (iii), the trajectory forms a closed periodic orbit, whose winding number per half cycle $T/2=\pi/\Omega$ is an integer $\alpha=1,2,3$, respectively.  
 }
\label{fig:classical}
\end{figure}
%%%%%%%%%%%%%%%%%%%%%%%%%%%%%%%%%%%%%%%%%%%%

Fig. \ref{fig:classical}(a) shows the results for the static diagonal and transverse conductivity  (all in absolute values) $\sigma^{0,0}_{yy}, \sigma^{0,0}_{xy}$ and the inverse heterodyne Hall conductivity  $1/\sigma^{1,0}_{xy}$, $1/\sigma^{0,1}_{xy}$ as a function of $r$. The diagonal conductivity $\sigma^{0,0}_{yy}$ first decreases when enlarging $B$ and vanishes at a discrete set of points,  labeled as  $r=r_{\alpha}^{\rm cl}$ ($\alpha=1,2,3,\ldots$).\\
This behavior can be understood from the dynamics of the particles in zero electric field  [Fig. \ref{fig:classical}(b)], which is no longer the cyclotron motion in an oscillating magnetic field.
In static but spatially inhomogenous fields, particles make detours and their paths were called ``snake states" \cite{PhysRevLett.68.385}. 
This generally also takes place  in temporary oscillating magnetic fields \cite{clas_traj} 
and the detour makes the particles ``heavy". 
We can relate the diagonal conductivity with the particle's effective mass $m_{\rm cl}^*$ by $m_e/m_{\rm cl}^*=\sigma_{yy}^{0,0}/\sigma_0$, with $\sigma_0= e^2n_e/(\eta m_e)$  being the zero field expression\cite{1999G}. 
When the diagonal conductivity $\sigma^{0,0}_{yy}$ vanishes  at  $r=r_{\alpha}^{\rm cl}$, the particle's trajectory in zero external field forms closed loops and the index $\alpha$ used to identify them has a topological meaning of winding number per half period. Indeed to have a closed  trajectory, no dissipative process should be present, which implies a vanishing diagonal conductivity.  The static transverse conductivity is expected to vanish due to time reversal invariance of the system on time scales multiples of a period.\\
The system also shows a nontrivial heterodyne Hall response. 
The Hall conductivity  $\sigma_{xy}^{0,1}$, which coincides with $\sigma_{xy}^{1,0}$,
takes values close to the classical result $ (n_e |e|c)/B$ when the field is strong enough [Fig. \ref{fig:classical}(a)]. 
In particular, they  coincide  when the effective mass diverges at $r=r_{\alpha}^{\rm cl}$; we note that this  feature is present also in the quantum case. 

%%%%%%%%%%%%%%%%%%%%%%%%%%%%%%%%%%%%%%%%%%%%
\section{Quantum case}\label{tre}
%%%%%%%%%%%%%%%%%%%%%%%%%%%%%%%%%%%%%%%%%%%%
Let us now consider a quantum version of the heterodyne Hall effect in a one-particle system. This is obtained with the minimal substitution starting from  free electrons ($p^\mu \to p^\mu -e/c A^\mu$, with $\mu=1,2$);  in the Landau Gauge 
the vector potential is $A_x(t)=0; A_y(t)=B_z(t)x$, which generates the electro-magnetic field
\begin{equation}
B_z(t)=B \cos\Omega t, \quad E_y(t)=-\frac{\partial B_z(t)}{\partial t}x.
\label{eq:BE}
\end{equation}
The  quantum Hamiltonian is
\begin{equation}
H(t)=\frac{\hbar^2 k_y^2}{2m_e}+H_0(x,t)-F(t)x
\end{equation}
where  $H_0(x,t)=\frac{p_x^2}{2m_e}+\frac{m_e(\omega(t))^2}{2}x^2$ is the Hamiltonian of a quantum harmonic oscillator (HO) with an oscillating frequency $\omega(t)=\omega_c\cos\Omega t$. $F(t)$ is a driving term which contains the (infinitesimal) input electric field, which we choose as $E_x(t)=E_x^1 \cos \Omega t$, and has the form $F(t)=\omega(t)\hbar k_y-e E_x(t)$.   We emphasize that translational invariance in the  $y$-direction still holds.

Using the time periodicity of the Hamiltonian $H(t+T)=H(t)$ for $T=2\pi/\Omega$, 
we seek for a solution of the time dependent Schr\"odinger equation in the  Floquet form  \cite{PhysRevA.7.2203,PhysRev.138.B979,grifoni1998driven,0022-3700-11-14-020}
\begin{equation}
\Psi_{n}(\Vect{x},t)=e^{-\frac{i}{\hbar}E_n(k_y) t}\Phi_n(\Vect{x},t),
\label{eq:solutionFloquet}
\end{equation} 
where $\Phi_n(\Vect{x},t)$ is a periodic function in time and $E_n(k_y)$ is the Floquet quasi-energy.  To be more precise, being (\ref{eq:solutionFloquet}) a Floquet solution, it will be labeled by a combined index $\alpha=(n,m)$ where $n$ is the HO energy level and $m=0,\pm 1,\pm 2,\ldots$ represents replica states (``photon-absorbed state'').
%We remove this redundancy by choosing the $m=0$ states
% to be in the 1st Brillouin Zone, i.e., $-\hbar \Omega/2<E_{(n,0)}\le  \hbar \Omega/2 $ and
% $E_\alpha=E_{(n,m)}=E_{(n,0)}+m\hbar \Omega$. 
 
Using the transformation by T. Taniuti and K. Husimi \cite{Husimi53}, 
the Floquet state \cite{PhysRevLett.66.527} is given by
\begin{eqnarray}
&&\Phi_{n}(\Vect{x},t)=\frac{e^{ik_yy}}{\sqrt{L_y}}\vphi_n(x-X(t),t)
\label{eq:solution}
\\
&&\times\exp\left[\frac{i}{\hbar }
\{m_e\dot{X}(t)(x-X(t))+\int_0^t dt'L(t')-L_0t\}\right].
\nonumber
\end{eqnarray} 
Here,  $\vphi_n(x,t)$ is the solution of the eigenvalue problem $\left[H_0(x,t)-i\hbar\frac{\pa}{\pa t}\right]\vphi_n(x,t)=\ve_n\vphi_n(x,t)$, 
 with energy $\ve_n$; The wavepacket center $X(t)$ is the solution of the  equation of motion for a classical HO  with a driving term $F(t)$: $m_e\ddot{X}(t)+m_e\omega(t)^2X(t)=F(t)$; $L(t)$ and $L_0$  are respectively the Lagrangian and its time average for this driven HO, given  by $L(t)=\frac{1}{2}m_e\dot{X}^2(t)-\frac{1}{2}m_e\omega(t)^2X^2(t)+X(t) F(t)$ and $L_0=\frac{1}{T}\int_0^{T}L(t')dt'$. The expression for $E_n(k_y)$ in (\ref{eq:solutionFloquet}) is related to $ \ve_n$ by
 \begin{equation}\label{tot_qe}
E_n(k_y)=\epsilon_n-L_0+ \frac{\hbar^2 k_y^2}{2m_e}.
\end{equation}
 
 %%%%%%%%%%%%%%%%%%%%%%%%%%%%%%%%%%%%%%%%%%%%% FIGURA PARTICLE PATHS
\begin{figure}[htb]
\centering 
\includegraphics[width=.4\textwidth]{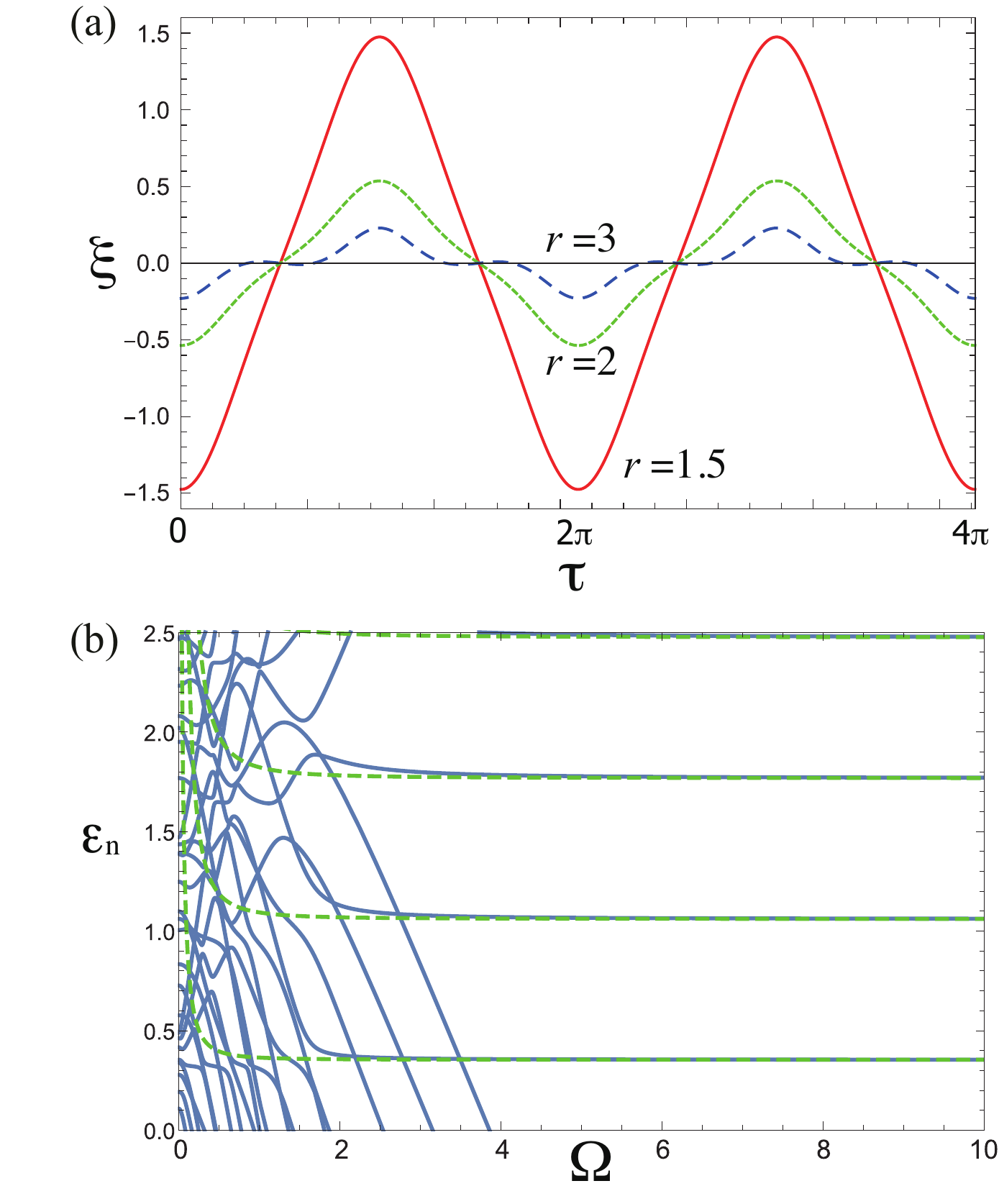}
\caption{(Color online) (a) Plot of $\xi(\tau)$  defined in (\ref{solX}) for $r=1.5,\;2,\;3$. 
(b) 
The intricate Floquet spectrum  (solid lines) reduces, for large enough $\Omega$, to equispaced   energy levels of a quantum HO (dashed lines) with frequency $\omega_{\rm eff}$.
}
\label{fig:pp}
\end{figure}
 %%%%%%%%%%%%%%%%%%%%%%%%%%%%%%%%%%%%%%%%%%%%% 

To extract the solution for $X(t)$ it is convenient to introduce a dimensionless variable $\xi(\tau)$ 
\begin{eqnarray}\label{solX}
X(t)=-(l_Br)^2\xi(\tau)\left(k_y-\frac{e E_x^1}{\hbar\omega_c }\right),
\end{eqnarray}
where $\tau=\Omega t$,  $l_B=\sqrt{\hbar c/eB}$ is the magnetic length and  $\xi(\tau)$ is the solution for Mathieu's equation with a source term 
%$\xi''(\tau)+(a-2q\cos 2\tau)\xi(\tau)=-\cos \tau$, 
$\xi''(\tau)+2a\cos^2\tau\xi(\tau)=-\cos \tau$, 
with $a=r^2/2$. The variable $\xi$ inherits the periodicity $\xi(\tau+2\pi)=\xi(\tau)$ from $X$ and oscillates around $x=0$ as shown in  Fig. \ref{fig:pp}(a).
This is obtained by writing $\xi(\tau)=\sum_m\xi_me^{-im\tau}$ and solving the linear relation $\xi_m(-m^2+\frac{r^2}{2})+\frac{r^2}{4}(\xi_{m-2}+\xi_{m+2})+\frac{1}{2}\delta_{m^2,1}=0$. 

To derive the Floquet spectrum  $\epsilon_n$ and the wave function $\vphi_n(x,t)$, we have to compute the Floquet Hamiltonian  $\mathcal{H}(x,t)=H_0(x,t)-i\hbar\frac{\pa}{\pa t}$, whose matrix elements in the Floquet basis $|m(t)\rangle= e^{-i m t}$ are given, as usual,  by
\begin{equation}\label{e1}
\mathcal{H}^{m,m'}=\frac{1}{T}\int dt e^{i(m-m')\Omega t} H_0(x,t)+ m \delta_{m,m'} \Omega. 
\end{equation}
After  conveniently rewriting $H_0(x,t)$  as 
\begin{equation}
H_0(x,t)=\frac{p_x^2}{2m_e}+\frac{m_e\bar{\omega}^2}{2}x^2+\frac{1}{4}m_e\bar{\omega}^2x^2 (e^{2 i \Omega t}+e^{-2 i \Omega t}),
\end{equation}
with $\bar{\omega}=\omega_c/\sqrt{2}$,  (\ref{e1}) yields
\begin{eqnarray}\label{full_eige}
\mathcal{H}^{m,m'}&=&\left(\frac{p_x^2}{2m_e}+\frac{m_e\bar{\omega}^2}{2}x^2 + m\hbar \Omega\right)\delta_{m,m'}\nonumber\\
&& +\frac{1}{4}m_e\bar{\omega}^2x^2(\delta_{m,m'+2}+ \delta_{m,m'-2}).
\end{eqnarray}
Using the Floquet-Magnus expansion \cite{effMag1, effMag2} and  $[[\frac{p^2}{2m},V(x)],V(x)]=-\frac{1}{m}\left(\frac{\pa V}{\pa x}\right)^2$, the high frequency effective Hamiltonian, up to order $\Omega^{-2}$, is
\begin{eqnarray}
H_{\rm eff}&=&H^{0,0}+\frac{[[H^{2,0},H^{00}],H^{2,0}]}{(2\Omega)^2}\nonumber\\
%+\mathcal{O}\left(\frac{1}{\Omega^4}\right)\nonumber\\
&=&\hbar\omega_{\rm eff}(\Omega)(n+1/2)
\end{eqnarray}
with $\omega_{\rm eff}(\Omega)=\frac{\omega_c}{\sqrt{2}}\sqrt{1+\frac{1}{16}\left(\frac{\omega_c}{\Omega}\right)^2}$. Therefore, in the large $\Omega$ limit, the energy eigenvalues reduce to those of a static quantum HO with  a renormalized frequency that depends on the driving $\Omega$
\begin{equation}\label{eige_eff}
\epsilon_n=\hbar\omega_{\rm eff}(\Omega)(n+1/2).
\end{equation}

In  Fig. \ref{fig:pp}(b) we present the full Floquet spectrum $\epsilon_n$ as a function of $\Omega$ (solid lines), obtained diagonalizing (\ref{full_eige}).    For large $\Omega$ values, it agrees well with the high frequency effective spectrum (\ref{eige_eff}) (dashed lines).
In this limit, the wave function $\vphi_n$ in (\ref{eq:solution}) is the usual HO eigenstate, i.e. $e^{-x^2/2l^2}H_n(x/l)$ with $l=(\hbar m_e/\omega_{\rm eff})^{1/2}$, 
while the orbital mixing increases as $\Omega$ becomes smaller.

\begin{figure}[htb]
\centering 
\includegraphics[width=.4\textwidth]{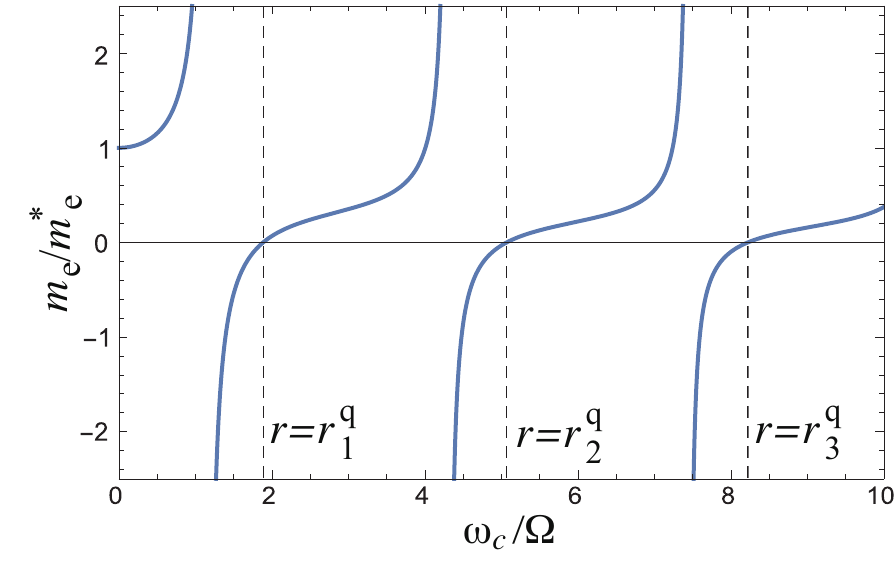}
\caption{(Color online) The inverse effective mass $\frac{m_e}{m^*_e}$ as a function of $r$. The effective mass diverges at $r=r_\alpha^{\rm q}$ resulting in the Landau quantization. 
}
\label{fig:5}
\end{figure}

%%%%%%%%%%%%%%%%%%%%%%%%%%%%%%%%%%%%%%%%%%%%
%{\it Landau quantization and heterodyne Hall effect:}
%%%%%%%%%%%%%%%%%%%%%%%%%%%%%%%%%%%%%%%%%%%%
The Floquet quasi-energy $E_n(k_y)$  can be further calculated from (\ref{tot_qe}) resulting in
\begin{equation}\label{moddisprel} 
E_n(k_y)=\epsilon_n+ \frac{\hbar^2 k_y^2}{2m_e}-\frac{\hbar^2}{2 m_e}\left(1-\frac{m_e}{m_e^*}\right)\left( k_y- \frac{eE_x^1}{\hbar \omega_c}\right)^2,
\end{equation}
where  the effective mass $m_e^*$ is given by
\begin{equation}
\frac{m_e}{m^*_e}=1+\frac{r^2}{2 \pi}\int_0^{2\pi}d\tau \cos \tau \xi(\tau)
\end{equation}
plotted in Fig. \ref{fig:5}. 
We can compare this plot with that of the classical longitudinal conductivity $\sigma_{yy}^{0,0}\sim m_e/m^*_{\rm cl}$  shown in Fig. \ref{fig:classical}(a). 
First, we see that the effective mass $m_e^*$ diverges at certain ratios $r=r_\alpha^{\rm q}$ collected in Table  \ref{table:T1}. 
When this happens, the $k_y$-dependence in (\ref{moddisprel}) drops out in the absence of $E_x^1$ and a macroscopic number of states become degenerate, 
an analog of  Landau quantization now realized by the oscillating magnetic/electric fields (\ref{eq:BE}). 
Around $r=r_\alpha^{\rm q}$, the effective mass changes sign from negative (hole like) to positive (electron like). 
Mathematically,  the condition for divergent ($r=r_\alpha^{\rm q}$) and zero effective mass, i.e.  $|m_e^*|\to \infty,0$, coincides with the periodic solution condition of the Mathieu equation without the source term.

%%%%%%%%%%%%%%%%%%%%%%%%%%%   TABLE    %%%%%%%%%%%%%%%%%%
\begin{table}
\begin{center}
\begin{tabular}{|p{1cm}||*{4}{c|}}
 %    \hline
\hline
\bfseries   $\alpha$  & 1 & 2 & 3 & 4\\
\hline
\bfseries $r_\alpha^{\rm q}$ &    1.89    & 5.07 & 8.22& 11.37 \\ 
\hline
\bfseries $Q$ &   0.221  & 0.153 & 0.124 & 0.106\\
\hline
\end{tabular}
\caption{The values for the constant $Q$ are reported for the first four $r=r^q_\alpha$.}
\label{table:T1}
\end{center}
\end{table}
%%%%%%%%%%%%%%%%%%%%%%%%%%% %%%%%%%%%%%%%%%%%%%%%%%%%%% 

%%%%%%%%%%%%%%%%%%%%%%%%%%%%%%%%%%%%%%%%%%%%% FIGURA RATIOMASSES
%%%%%%%%%%%%%%%%%%%%%%%%%%%%%%%%%%%%%%%%%%%%
The quantum version of the heterodyne Hall effect occurs when we turn on the $x$-direction ac electric field $E_x^1$ leading to a dc current flowing in the $y$ direction.  
We can  compute the dc-current $J_y(k_y)=e\frac{\pa E_n}{\hbar\pa k_y}$ for a state with $k_y$ as the momentum derivative of the dispersion relation (\ref{moddisprel})
as in the static case. 
%\begin{equation}\label{Jy}
%J_y(k_y)=q\frac{\pa E_n}{\hbar\pa k_y}=\frac{q p_y}{m_e^*}+\frac{q^2}{m_e}\left(1-\frac{m_e}{m_e^*}\right)\frac{E_x^1}{\omega_c}.
%\end{equation}
The total current density for a system of dimensions  $L_x \times L_y$ is defined as $j_y=\frac{1}{L_x L_y}\sum_{k_y} f_n(k_y)J_y(k_y)$
and given that the distribution $f_n(k_y)$ is even in $k_y$ due to the  invariance under time reversal, we obtain a linear relation
\begin{equation}\label{eq:qhhe}
%j_y=\frac{ e^2 \nu N_\Phi}{L_x L_y \omega_c m_e}\left(1-\frac{m_e}{m_e^*}\right)E_x^1,
j_y=\sigma^{0,1}_{yx}E_x^1,
\end{equation}
where the heterodyne Hall coefficient is given by 
\begin{eqnarray}\label{eq:sigmaxyq}
\sigma^{0,1}_{yx}=\frac{e^2 }{h} Q\nu.
\end{eqnarray}
The Landau level filling $\nu=N_e/N_\Phi$ is defined as the ratio of the electron density $N_e$ and the level degeneracy 
\begin{equation}\label{dege}
N_\Phi=\frac{L_xL_y}{2\pi l_B^2r^2{\rm max}\;\xi};
\end{equation}
$N_\Phi$ is obtained by imposing the wave packet center (\ref{solX})
to be within the strip, i.e. $X(t)\in [-L_x/2,L_x/2]$ for $E_x^1=0$, where ${\rm max}\;\xi$
is the maximum of $\xi$ during time evolution. 
The factor $Q={\left(1-\frac{m_e}{m^*_e}\right)}/{ \left(2r^2{\rm max}\xi\right)}$ is a nonmonotonous function of $r$,
while its value at $r=r_{\alpha}^{\rm q}$ presented in Table~\ref{table:T1} monotonously decreases. 

%%%%%%%%%%%%%%%%%%%%%%%%%%%%%%%%%%%%%%%%%%%%% FIGURA Toro
\begin{figure}[tbh]
\centering 
\includegraphics[width=.45\textwidth]{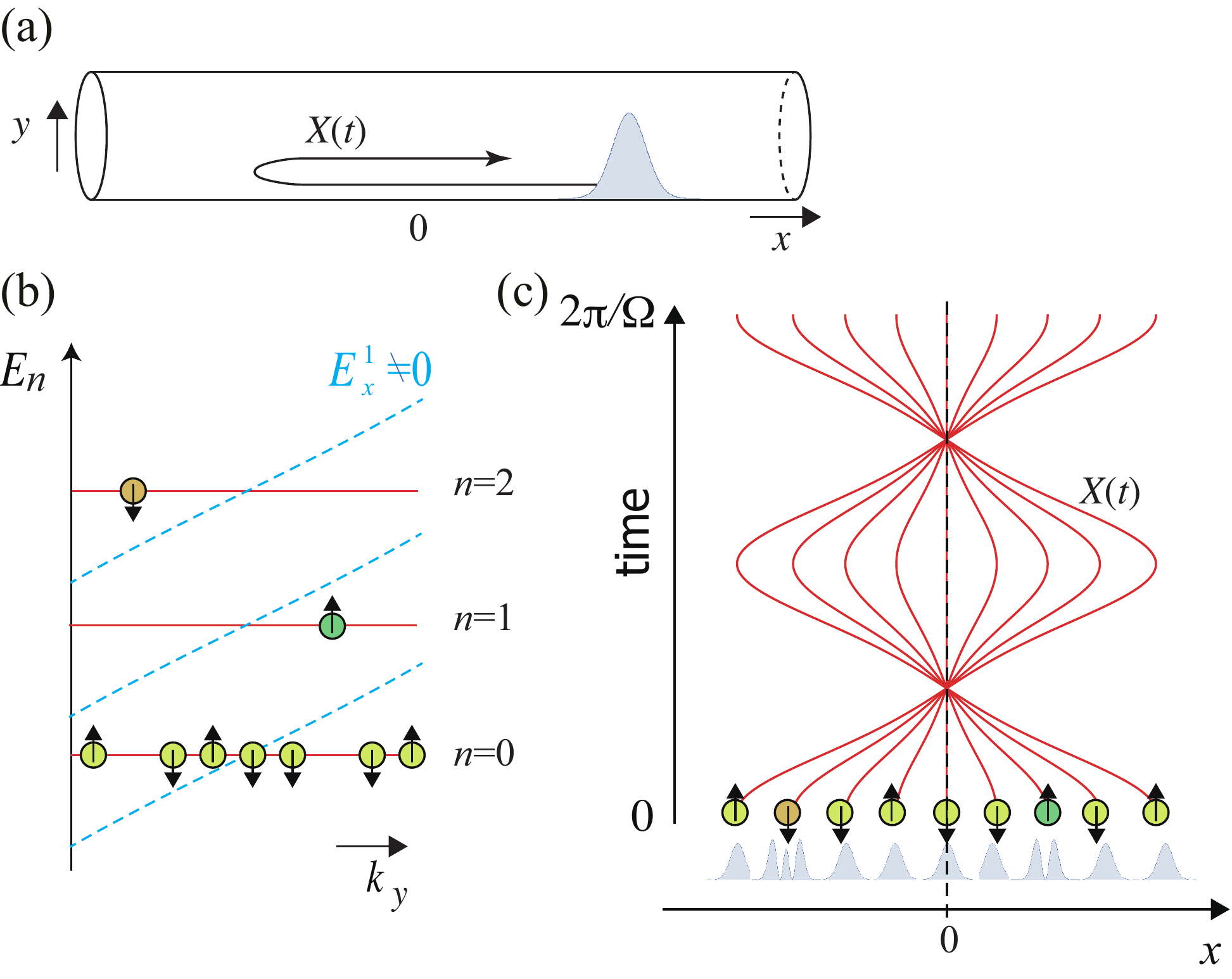}
\caption{(Color online) (a) Schematic representation of the particle's dynamics. The wavefunction is a plane wave with momentum $k_y$ along the $y$-direction (represented as a compactified dimension, assuming $L_y$ finite)  and a localized wave packet  oscillating along the $x$-direction. 
%The initial position $X(0)$ depends on $k_y$; wavepackets with different $k_y$ evolve independently and oscillate around $x=0$ according to (\ref{solX}). 
(b) When the Landau quantization condition  $r=r_{\alpha}^{\rm q}$
is met, the Floquet quasi-energy becomes flat (linearly tilted) 
in the absence (presence) of %the extra $x$-direction ac electric field of the form 
$E_x(t)=E_x^1 \cos(\Omega t)$.
The many particle state is achieved by filling the states with electrons with a spin, denoted as circles with an arrow, respecting the Pauli principle. 
(c) The motion of the wave packet center for the many particle state in (b). 
The initial position $X(0)$ depends on $k_y$; wavepackets with different $k_y$ evolve independently and oscillate around $x=0$ according to (\ref{solX}). 
%States with different $k_y$ and Landau level index $n$ oscillates around $x=0$. 
}
\label{fig:metafig}
\end{figure}
%%%%%%%%%%%%%%%%%%%%%%%%%%%%%%%%%%%%%%%%%%%% 

\section{Conclusions}\label{quattro}
To summarize, we have computed the heterodyne conductivities in a 2DEG subject to a time oscillating magnetic field, both for the classical and quantum case.

We schematically illustrate our findings in Fig. \ref{fig:metafig} and discuss several problems we would like to investigate in the future. 
(i) The many particle state is realized by filling the states with $N_e=\nu N_\Phi$ electrons as indicated in Fig. \ref{fig:metafig}(b). 
Since the system is heated by the external driving, it is likely to have states with mixed Landau orbitals $n$. 
The effect of Coulomb interaction may lead to interesting effects. The electron wave functions overlap simultaneously around $x=0$ during the 
time evolution [Fig. \ref{fig:metafig}(c)]. This makes the interaction between states with different $k_y$ to be enhanced and long-ranged. If the system 
can be stabilized and cooled, ordering such as ferromagnetism, Wigner crystal, and even an analogue of the fractional QHE state might be induced. 
However,  it is also likely that the accumulation of macroscopic number of electrons will make the system unstable
and even  destroy the sample along the line $x=0$. 
(ii) Is the heterodyne Hall conductivity $\sigma_{yx}^{0,1}$ a topological quantity? 
Similar to the traditional  IQHE \cite{Klitzing80,1999G}, the current expression (\ref{eq:sigmaxyq}) is proportional to $\nu$ and is thus quantized. 
The renormalized coefficient $\frac{e^2 }{h} Q$ is fixed as long as the magnetic field $B$ is changed 
simultaneously  with the frequency $\Omega$ respecting the quantization condition $r=r_{\alpha}^{\rm q}$. 
In order to answer this question, an edge calculation and an extension of the TKNN formula\cite{TKNN} is important, 
which may reveal a bulk-boundary correspondence \cite{Hatsugai93} in heterodynes. 
(iii) Physical realization is an important problem. 
The driving field (\ref{eq:BE}) can be realized by placing two anti-parallel wires 
with currents oscillating as $\pm I\cos (\Omega t)$. 
The 2DEG is to be placed between the wires. 
This setup may be realized using THz plasmonics, with which it is already possible to generate magnetic fields with strength above 
1 Tesla oscillating in the terahertz domain\cite{Yen04,Mukai}. This is the strength and 
frequency necessary to realize the quantization condition and to be in the 
quantum limit, {\it i.e.} small $\nu$.

\section{Acknowledgments} 
We thank Masaaki Nakamura and  Yu Mukai  for illuminating discussions. 
TO acknowledges Stefan Kaiser, Thomas Weiss, Koichiro Tanaka and Andre Eckardt for fruitfull discussions.  
This work is partially supported by KAKENHI (Grant No. 23740260) and from the ImPact project (No. 2015-PM12-05-01) from JST. 

%bibliography

\bibliographystyle{revtex.bst}
\bibliography{PRB_heterodyne_v1.bib}
%\printindex

\end{document}